\documentclass[1p]{elsarticle}

\usepackage{lineno,hyperref,natbib}
\modulolinenumbers[1]

\usepackage{xr}
\usepackage{bm}
\usepackage{soul}
\usepackage{amsmath,amssymb,latexsym}
\usepackage{booktabs} 
\usepackage[table]{xcolor} 
\usepackage{array} 
\usepackage{rotating} 
\newcolumntype{P}[1]{>{\centering\arraybackslash}p{#1}}

\usepackage{graphics,dcolumn,epic, eepic,fleqn,float}
\usepackage{amsfonts,multirow,rotate,color}
\usepackage{wasysym}
\usepackage{hyperref}

\biboptions{sort&compress}

\begin{document}

\begin{frontmatter}

\title{Network neuroscience for optimizing brain-computer interfaces}

\author[1,2,3,4,5]{Fabrizio De Vico Fallani}
\author[6,7,8,9]{Danielle S. Bassett}


\address[1]{Inria Paris, Aramis project-team, F-75012, Paris, France}
\address[2]{Institut du Cerveau et de la Moelle ́Epiniere (ICM), F-75013, Paris, France}
\address[3]{Inserm, U 1127,  F-75013, Paris, France}
\address[4]{CNRS, UMR 7225, F-75013, Paris, France}
\address[5]{Sorbonne Universite, F-75013, Paris, France}
\address[6]{Department of Bioengineering, University of Pennsylvania, Philadelphia, PA, 19104, USA}
\address[7]{Department of Electrical and Systems Engineering, University of Pennsylvania, Philadelphia, PA, 19104, USA}
\address[8]{Department of Physics and Astronomy, University of Pennsylvania, Philadelphia, PA, 19104, USA}
\address[9]{Department of Neurology, Perelman School of Medicine, University of Pennsylvania, Philadelphia, PA, 19104, USA}

\begin{abstract}
Human-machine interactions are being increasingly explored to create alternative ways of communication and to improve our daily life. Based on a classification of the user's intention from the user's underlying neural activity, brain-computer interfaces (BCIs) allow direct interactions with the external environment while bypassing the traditional effector of the musculoskeletal system. Despite the enormous potential of BCIs, there are still a number of challenges that limit their societal impact, ranging from the correct decoding of a human's thoughts, to the application of effective learning strategies. Despite several important engineering advances, the basic neuroscience behind these challenges remains poorly explored. Indeed, BCIs involve complex dynamic changes related to neural plasticity at a diverse range of spatiotemporal scales. One promising antidote to this complexity lies in network science, which provides a natural language in which to model the organizational principles of brain architecture and function as manifest in its interconnectivity. Here, we briefly review the main limitations currently affecting BCIs, and we offer our perspective on how they can be addressed by means of network theoretic approaches. We posit that the emerging field of network neuroscience will prove to be an effective tool to unlock human-machine interactions.
\end{abstract}

\begin{keyword}
\texttt{Human-machine interaction \sep Neurofeedback \sep  Neural plasticity \sep Brain connectivity \sep Graph theory}
\end{keyword}

\end{frontmatter}


\newpage
\section{Perspective}
Brain-computer interfaces (BCIs) have been developed to translate brain activity into informative signals that can be used by external devices. BCIs allow a direct interaction between humans and machines, and are increasingly used for control and communication, as well as for the treatment of neurological disorders \citep{wolpaw_braincomputer_2002, daly_braincomputer_2008}. Since the first proof-of-concept studies demonstrating the possibility to move a graphical object on a computer screen by means of electroencephalography (EEG)\citep{vidal_toward_1973}, research developments in this area have increased exponentially \citep{schalk_bci2000:_2004,muller-putz_eeg-based_2005,schwartz_brain-controlled_2006,cincotti_high-resolution_2008,carlson_brain-controlled_2013,lafleur_quadcopter_2013,pichiorri_braincomputer_2015}. 

BCIs hold tremendous potential for open-loop (control) and closed-loop (biofeedback) applications, particularly via their ability to exploit subjects' voluntary control over their brain activity through mental imagery (MI). Despite this potential, the societal and clinical impact of BCIs has so far been rather limited due to their poor reliability in the user’s daily life \citep{clerc_brain-computer_2016}. Indeed, BCI performance as measured by accurate classification of the user's intent, is still relatively variable and does not provide the guarantees of functioning that are necessary in most clinical scenarios. One of the greatest challenges is to understand and solve the problem of ``BCI illiteracy'', which refers to a phenomenon that occurs in a non-negligible portion of users (estimated to be around 15-30$\%$) who are not able to properly use a BCI \citep{vidaurre_towards_2010}.

While many solutions have been proposed -- from the identification of the best mental strategy to code the user’s intent, to the optimization of brain features, type of sensory feedback, and classification algorithm -- the results are still not satisfactory and more research is needed \citep{bougrain_brain-computer_2016}. Here, we focus on the fundamental role of brain features as the substrate for the BCI algorithm. There are basically two antithetic approaches widely adopted in the literature. The first one extracts features from the activity of specific brain sites that are related to the mental strategy. This approach is, for example, the one used in motor imagery-based BCIs, where power spectra in the primary motor areas is the chosen feature \citep{pfurtscheller_event-related_1999}. The second approach instead takes into account all of the available information by computing, for example, the covariance matrix of all sensor signals. While the latter approach is particularly suitable for advanced classification algorithms \citep{barachant_multiclass_2012}, it hampers the simple identification of underlying neurophysiological mechanisms. 

Here we offer a complementary perspective that ideally combines the advantages of the previous approaches. We begin by acknowledging that one cannot infer neural mechanisms from a collection of disconnected parts, but instead must obtain an understanding of the system's collective behavior. Examining the activity of one specific region -- while neglecting its interactions with other regions -- oversimplifies the true phenomenon. To embrace the substratal complexity, we consider the human brain as a complex network where regions are both anatomically and functionally wired together with one another. Network science provides a natural language to describe such networks by modeling them as mathematical objects called graphs \citep{albert_statistical_2002, newman_structure_2003, boccaletti_complex_2006}.

One of the advantages of the network approach is the ability to extract summary statistics or metrics that quantitatively measure specific organizational characteristics across a variety of topological scales. Network metrics have been used to demonstrate, for example, that brain networks exhibit modular structure, where groups of brain regions display highly clustered connectivity at the mesoscale. Regions within modules tend to interact preferentially through short-distance links \citep{betzel2017modular}, while regions across modules tend to interact through highly connected nodes known as hubs \citep{bertolero2015modular}. While these organizational properties support basic cognitive functions, such as a balance between integration and segregation of information, they are nevertheless sensitive to pathological and physiological alterations of the mental state \citep{bullmore_complex_2009, park_structural_2013,stam_modern_2014}.
The connection between network topology and function underscores the potential for using network science as an effective tool for improving BCI performance.

\begin{figure}[!ht]
\centering
\includegraphics[width=\textwidth]{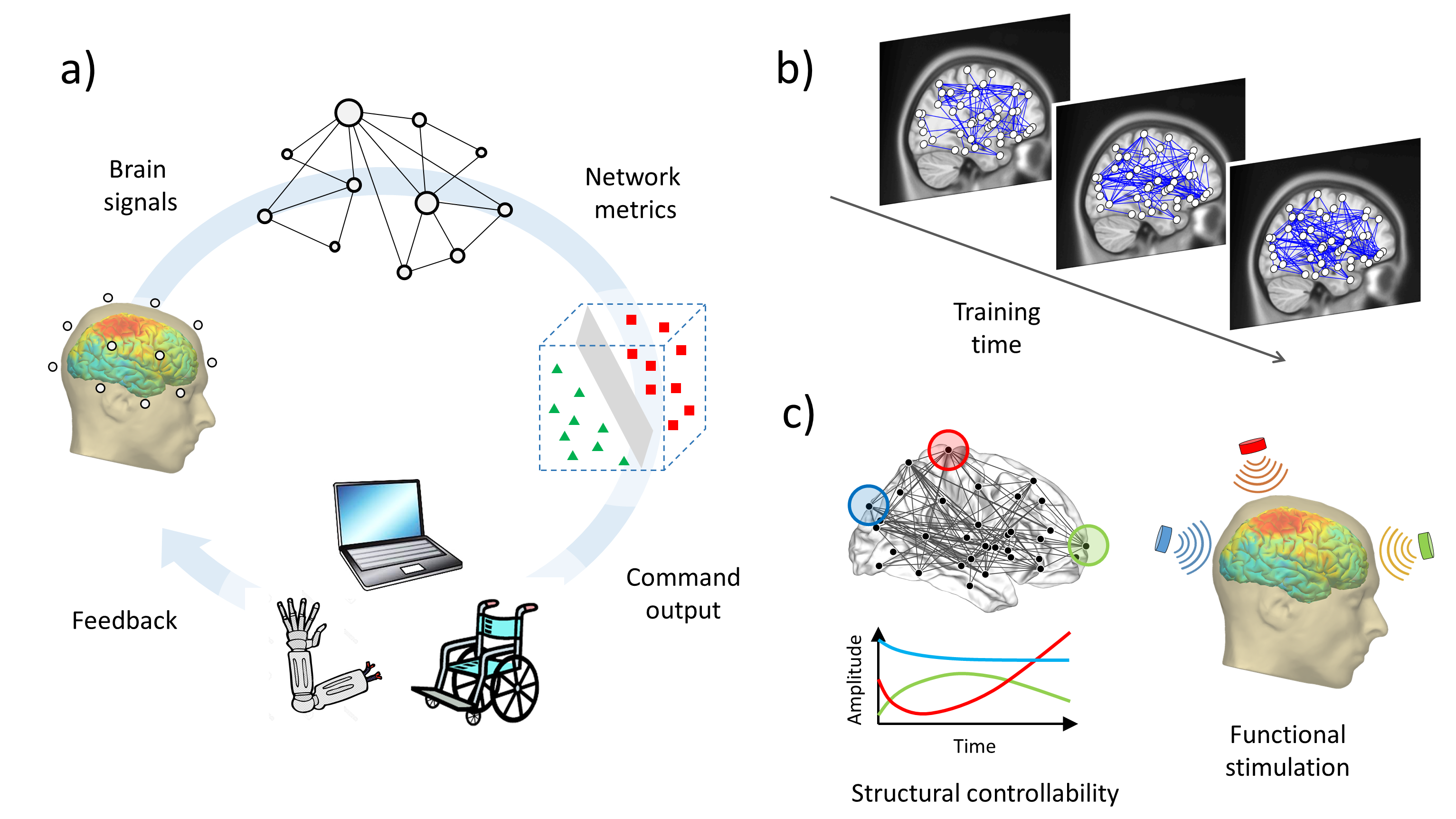}
\caption{\textbf{A network neuroscience approach to brain-computer interfaces (BCIs).} \textit{Panel a)} Network-based brain-computer interfaces. The user modulates her or his brain activity to control the BCI. Brain signals are recorded through sensors such as electro/magnetoencephalography (E/MEG). Functional connectivity is used to infer the corresponding interaction network or \emph{graph}. Different network metrics are extracted to constitute the feature vector (i.e., a point in the scatter plot). Machine learning algorithms use this feature vector to classify the user's mental states (i.e., the red squares and green triangles in the scatter plot) generated during the experiment. At each time point, the final result is sent to the external device that executes the command and gives the feedback to the user. \textit{Panel b)} Quantification of neural plasticity during BCI training. Temporal network metrics, which describe higher-order time-varying connectivity changes, can be used to model dynamic brain networks obtained longitudinally from neuroimaging signals such as functional magnetic resonance imaging (fMRI). These metrics, reflecting transient organizational mechanisms, are suitable candidates to predict future BCI performance. \textit{Panel c)} Principles of brain network controllability for modulating function. Network control theory is used to identify the driver nodes in the structural connectome obtained from diffusion tensor imaging data (DTI), and also to derive the theoretically predicted signals needed to change the brain activity. Noninvasive functional brain stimulation, such as transcranial magnetic stimulation (TMS), can be then used to experimentally favor detectable brain activity patterns.}
    \label{fig:1}
\end{figure}

Network metrics can be used as complementary brain features in BCIs (Fig. \ref{fig:1}a). Recent studies have demonstrated their potential in discriminating between different mental states related to BCI tasks \citep{demuru_brain_2013, xu_motor_2014}. During BCI experiments, metrics must be computed from time-varying brain networks in order to give real-time feedback to the user \citep{khambhati2017modeling,sizemore2017dynamic}. The feasibility of tracking network metrics in real-time has already been demonstrated by using dynamic functional connectivity (dFC) measures from EEG \citep{de_vico_fallani_persistent_2008}, MEG \citep{valencia_dynamic_2008}, and fMRI \citep{calhoun_time-varying_2016} signals. However, the time required to compute some metrics (e.g., shortest path length) can become intractable when the number of nodes $N$ is large, i.e. $N>100$. Furthermore, the statistical reliability of the estimated functional connections significantly decreases with the length of the time window considered \citep{de_vico_fallani_graph_2014,preti_dynamic_2017}. While possible solutions are available based on efficient sparse-coding algorithms \citep{valdes-sosa_estimating_2005,lee_sparse_2011}, the statistical reliability of the estimates remains the main challenge for the effective use of network metrics online. Eventually, having reliable temporally dynamic brain networks will allow researchers to exploit the nascent formulation of (i) multilayer networks to extract temporal metrics, which can be used to quantify higher-order properties such as persistence or flexibility \citep{tang_small-world_2010,bassett_dynamic_2011}, as well as (ii) multiplex networks to extract metrics quantifying cross-frequency functional interactions \citep{de_domenico_mapping_2016,guillon_loss_2017}.

Once extracted, the different network metrics constitute the feature vector for the classification of the user's mental state. This inference of the mental state from the feature vector can then be used to transmit the correct command to the computer (Fig. \ref{fig:1}a). Interestingly, brain networks are a particular case of graphs where nodes correspond to specific spatial sites (i.e. the brain areas) and only their connections are allowed to change \citep{bassett2010efficient,de_vico_fallani_graph_2014}. This fact implies that the size of the feature vectors including nodal metrics - such as the degree - will not change across mental conditions and can be directly input to statistical machine-learning algorithms or to the mass-univariate tests in order to optimally perform the classification \citep{zanin_optimizing_2012,richiardi_machine_2013}. Notably, this same spatially-embedded property of the brain can further exploited to fine-tune statistical null models of the brain networks involved: both for the purposes of comparing the observed features to those anticipated in the null, and for the purposes of incorporating null expectations into the estimated features themselves \citep{samu2014influence,roberts2016contribution,betzel2017modular}.

Controlling a BCI is a learned skill based on the feedback presented to the user. In general, several weeks or even months are needed to obtain high performance, and in some cases adequate control is never reached \citep{vidaurre_towards_2010, jeunet_why_2016}. This gap in the ease with which different individuals learn to effectively use a BCI has motivated scientists to consider adaptive BCI architectures that can dynamically accommodate the transient nature of brain features \cite{shenoy_towards_2006,vidaurre_co-adaptive_2011}. In fact, during BCI skill acquisition, users often report transitioning from a deliberate cognitive strategy (e.g., motor imagery) to a nearly automatic goal-directed approach focused directly on effector control \citep{wander_distributed_2013}. This evidence is indicative of a network reconfiguration process that is consistent with procedural motor learning. Efforts to better understand the neural dynamics underlying BCI training have capitalized on a range of neuroimaging techniques in both humans and non-human primates \citep{carmena_learning_2003,jarosiewicz_functional_2008,wander_distributed_2013, toppi_investigating_2014,kaiser_cortical_2014}. Results have shown that even if BCIs typically receive inputs from a few brain regions, a distributed network of remote cortical areas is actually involved throughout BCI skill acquisition. 

Network neuroscience approaches have recently been adopted to quantify brain network reorganization underlying diverse types of human learning (Fig. \ref{fig:1}b). For example, network flexibility of association areas in fMRI-based functional brain networks, measured as the fraction of times that a node changed its allegiance to a functional module throughout training \citep{bassett_dynamic_2011}, has been showed to positively predict individual differences in motor learning, cognitive control, and executive function \citep{braun2015dynamic}. The same statistic, when calculated from the ventral striatum has also been shown to positively predict individual differences in accuracy on a reinforcement learning task, as well as on reinforcement learning rate parameters estimated from mathematical models of the individual's behavior \citep{gerraty2018dynamic}. Resting state markers of this flexible module architecture have noted capacity to predict future learning over 6 weeks of motor skill training \citep{mattar2018predicting}. Similarly, a decreased functional integration, measured by shortest paths between nodes in EEG-derived functional brain networks, has been reported after motor imagery training, and was interpreted as a putative marker of an underlying automaticity process \citep{pichiorri_sensorimotor_2011}. These results suggest that network science holds the potential to unveil the neural basis of BCI learning and predict future performance, thereby informing the optimization of adaptive BCI architectures. 

It is important to admit that while richer brain features and an enhanced understanding of the process of learning itself may enhance BCI performance on average, challenges may still remain for single individuals. Indeed, for some users, it may be impossible to rapidly generate an appropriate activity pattern that is accurately detected by the machine. An alternative approach is to draw on recent advances in neurostimulation technology, such as transcranial magnetic (TMS) or direct current (tCDS) stimulation,  which can directly influence brain state by altering network dynamics \citep{chen2013causal}. Such technology has notable potential, but immediate applications have been hampered by the lack of an understanding of how and where to stimulate to generate a desired mental state \citep{johnson2013neuromodulation}. Gaining this understanding will require informed models that can \emph{a priori} produce predictions about where and how to deliver stimulation to induce a specific pattern of brain activity. One recently proposed model builds on notions of network controllability \citep{gu2015controllability}, where stimulation is stipulated to pass along white matter tracts and therefore where stimulation-induced change in brain state is constrained by the structural connectome \citep{muldoon2016stimulation}. 

Initial applications of the theory of network control to neural systems has spanned a wide range of species, including \emph{C. elegans}, mouse, \emph{Drosophila}, macaque, and human, and has ranged from data through models to pure theory \citep{wiles2017autaptic,gu_controllability_2015,kim2018role,yan2017network}. Applications to the clinic have largely focused on questions of predicting and altering seizure dynamics, although recent work has demonstrated utility in the understanding of psychiatric disorders such as bipolar \citep{taylor2015optimal,ching2012distributed,ehrens2015closed,jeganathan2018fronto,braun2018maps}. Network control has been suggested to have utility in neurofeedback specifically and BCIs more generally \citep{bassett2017network,murphy2017network}, in part due to the marked correspondence between theoretically-predicted control points in the brain and the cognitive functions they support across development and in healthy adulthood \citep{gu2015controllability,tang2017developmental}. Future efforts further validating or extending the network control model may serve as an important complement to efforts in BCI feature selection and optimization.

To conclude, here we have provided our perspective on why and how network science has the potential to improve the performance of brain-machine interactions. By offering this perspective, we hope to stimulate a global and interdisciplinary discussion to collectively identify the elements of BCI learning that should be reconsidered, in an effort to boost their societal and clinical impact.

\section*{Acknowledgments}
FDVF acknowledges support by the ANR French program through the contracts ANR-10-IAIHU-06 and ANR-15-NEUC-0006-02. DSB acknowledges support from the John D. and Catherine T. MacArthur Foundation, the Alfred P. Sloan Foundation, the ISI Foundation, and the NIH 1R01HD086888-01. The funders had no role in study design, data collection and analysis, decision to publish, or preparation of the manuscript.


\bibliography{final}

\newpage

\end{document}